# An X-Band Waveguide Measurement Technique for the Accurate Characterization of Materials with Low Dielectric Loss Permittivity

Kenneth W. Allen,[1, a)] Mark M. Scott,[1, b)] David. R. Reid,[1] Jeffrey A. Bean,[1] Jeremy D. Ellis,[1] Andrew P. Morris,[1] and Jeramy M. Marsh[1]

[1] *Advanced Concepts Laboratory, Georgia Tech Research Institute, Atlanta, GA 30318 USA,*

**Footnotes:**
a)   kenneth.allen@gtri.gatech.edu
b)   Current address: Resonant Sciences LLC, Dayton, OH 45430 USA.

In this work, we present a new X-band waveguide (WR90) measurement method that permits the broadband characterization of the complex permittivity for low dielectric loss tangent material specimens with improved accuracy. An electrically-long polypropylene specimen that partially fills the cross-section is inserted into the waveguide and the transmitted scattering parameter ($S_{21}$) is measured. The extraction method relies on computational electromagnetic simulations, coupled with a genetic algorithm, to match the experimental $S_{21}$ measurement. The sensitivity of the technique to sample length was explored by simulating specimen lengths from 2.54 to 15.24 cm, in 2.54 cm increments. Analysis of our simulated data predicts the technique will have the sensitivity to measure loss tangent values on the order of $10^{-3}$ for materials such as polymers with relatively low real permittivity values. The ability to accurately characterize low-loss dielectric material specimens of polypropylene is demonstrated experimentally. The method was validated by excellent agreement with a free-space focused-beam system measurement of a polypropylene sheet. This technique provides the material measurement community with the ability to accurately extract material properties of low-loss material specimen over the entire X-band range. This technique could easily be extended to other frequency bands. © *2016 American Institute of Physics*. [DOI: XX.XXXX/X.XXXX]

## I. INTRODUCTION

Waveguide (WG) measurement techniques have been successfully implemented for the characterization of dispersive complex permittivity ($\varepsilon(\omega)^* = \varepsilon(\omega)' - j\varepsilon(\omega)''$) of dielectric material specimens for nearly seven-decades[1-6]. However, there have been significant recent improvements in these WG measurement techniques due to advances in computational electromagnetic (CEM) simulation tools and numerical non-linear solvers[7,8]. These computational tools have relaxed the geometrical restrictions on the specimen, allowing for the characterization of specimens with irregular shapes as well as complex anistropy[8]. Despite these advances, the most accurate methods for the RF/microwave permittivity characterization of low-loss dielectrics remain resonant methods[9-12]. These resonant methods are inherently restricted to discrete frequencies and particular specimen-WG geometries, as they rely on perturbation theory. Also, resonant cavities are challenging to model using CEM tools due to high quality factors.

The method presented in this work enables high accuracy broadband permittivity characterization of low dielectric loss tangent (tan$\delta$) material specimens. The high accuracy is facilitated by the use of electrically long material specimens to increase the field-sample interaction, separating this technique to previous studies that utilized the partially filed method for the extraction of low-loss materials[6]. Material properties are extracted through CEM-based analysis of the measured transmitted scattering parameters ($S_{21}$) of an X-band WG with an electrically long material specimen that partially fills (25%) the cross-section of the WR90 WG. A sensitivity analysis is performed to determine the necessary sample length by varying the length over the range from 2.54-15.24 cm. The measurement sensitivity is observed to be enhanced by increasing the electrical length of the specimen.

The complex permittivity extraction algorithm implements a finite element method (FEM) model that solves frequency-by-frequency the desired scattering parameters. As the complex permittivity of the unknown material under test is varied within this simulation, the simulated scattering parameters are evaluated by a scoring algorithm, or fitness function, which quantifies the difference between simulated and experimentally obtained results. The result of the fitness function, in turn, drives a genetic algorithm to modify the permittivity values of the material under test to improve the match between the simulated and experimental scattering parameters.

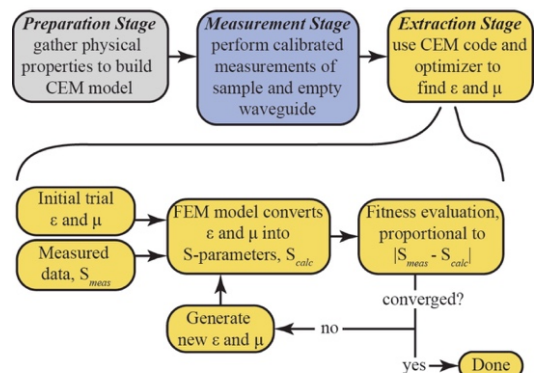

FIG. 1. Schematic illustration of the extraction methodology.

This process is schematically illustrated in Fig. 1. This evaluation is performed over a range of frequencies, permitting the complex permittivity extraction over the entire measured frequency band (e.g., 8.2-12.4 GHz). While not demonstrated in this work, this method allows for the extraction of high permittivity dielectrics and dispersive properties of the material specimen, which is not possible with resonant methods due to cavity over-perturbation and, in some cases, air gap errors. Additionally, the concept presented herein can be easily extended to other waveguide bands.

To demonstrate and validate the capability and accuracy of this method for low-loss specimens, a known low dielectric loss tangent material (polypropylene) was measured in the WR90 WG. The permittivity results were compared with values published in the literature and with results acquired from a free-space focused-beam system (FBS) measurement. The sensitivity of the phase and amplitude of $S_{21}$ for this method was investigated by simulating changes in $\varepsilon(\omega)'$ by ±1%, while maintaining $\tan\delta$, and also by simulating changes in $\varepsilon(\omega)''$ by factors of ½ and 2 of 0.003 while maintaining a constant real permittivity. The sensitivity of the method is compared with the measurement repeatability of the network analyzer.

## II. EXTRACTION METHODOLOGY

A three-stage process is implemented in order to accurately extract the complex permittivity values for the material specimen, as illustrated in Fig. 1: 1) a preparation stage, 2) a measurement stage, and 3) an extraction stage. This three-stage process is delineated in more detail in previous work[8]. It is also important to note, however, that previously the extraction process was performed by examining the reflected scattering parameter ($S_{11}$) for multiple measurements of an anisotropic material specimen. To accurately characterize the loss tangent of low-loss materials, however, the technique presented in this work relies on a single $S_{21}$ measurement performed on a material specimen that is presumed to be isotropic. This method can be extended for anisotropic materials by taking additional measurements in order to obtain $S_{21}$ values for different axis orientations along the propagation direction.

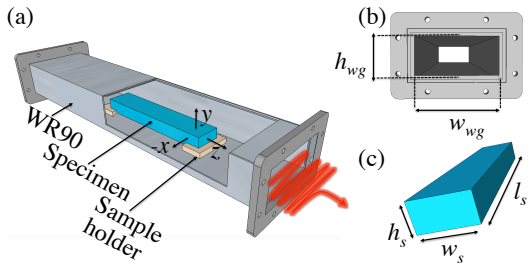

FIG. 2. Schematic illustrations of (a) specimen partially filling the cross-section of the X-band waveguide, (b) labeled dimensions of the waveguides cross-section, and (c) specimens dimensions labeled.

In the preparation stage, the WR90 WG dimensions ($w_{wg}$=2.286 cm, $h_{wg}$=1.143 cm, $l_{wg}$=35.56 cm) and material specimen dimensions ($w_s$=1.176 cm, $h_s$=0.521 cm, $l_s$=12.675 cm) are collected as schematically illustrated in Fig. 2. Nominally, the material specimen was designed such that $w_s = w_{wg}/2$, $h_s = h_{wg}/2$, and $l_s = 12.7$ cm; however, due to fabrication tolerances, the specimen cross-section deviated from the nominal design. The actual specimen dimensions are an integral component to the accuracy of the extraction process. Deviations of the specimen dimensions are typically on the order of a millimeter, and this is considered and incorporated into the final stage of the extraction process to ensure an accurate model geometry. This illustrates one of the significant advantages of this technique. It is frequently the case that specimens can be characterized with greater precision than they can be machined. Unlike traditional, fully-filled waveguide techniques, machining imperfections do not limit the accuracy of this technique, as long as the 'as machined' dimensions can be incorporated into the CEM model. The final component in the preparation stage is performing a thru-reflect-line (TRL) calibration on the WR90 WG measurement fixture.

In the measurement stage, $S_{21}$ measurements are performed with and without the material specimen inserted into the WR90 WG. Empty WG measurements are considered as an additional fixture compensation such that $S_{21} = S_{21}^{sample}/S_{21}^{empty}$, where $S_{21}^{sample}$ and $S_{21}^{empty}$ are the measurements of the transmitted scattering parameters with and without the material specimen, respectively. Note that, during the measurement of $S_{21}^{empty}$, the low-density foam sample holders are placed in the WG. This allows the calibration to account for the reduced amplitude and phased delay accrued due to the sample holder, although the effect is generally negligible. The aforementioned calibration and compensation of $S_{21}$ is performed, in part, to ensure that the phase reference planes align for the measured and simulated scattering parameters. Additionally, the use of this fixture compensation term helps to mitigate artificial numerical grid dispersion that is an inherent source of error in the the simulated scattering parameters.

In the final stage of the characterization process, the extraction stage, iterative computations are performed using the experimentally obtained $S_{21}$ as the target value, which allows for the complex permittivity values of the material under test are varied in the search space by the numerical solver. The amount of time required to extract material properties depends directly on the efficiency of the FEM model, in terms of mesh and volume. Often, thousands of simulations are performed per frequency in order to fit the simulated and experimental $S_{21}$. For this work, properties were extracted at 30 frequencies within the band of interest; however, this is arbitrary in the sense that more or less can be used as desired. Prior to beginning the extraction process, therefore, convergence studies were performed in order to ensure that the FEM model geometry was meshed in a manner that provided the required accuracy while keeping runtime at a minimum. The convergence study resulted in an adaptive tetrahedral mesh, shown in Fig. 3, with the maximum mesh elements inside the specimen volume meshed at $\lambda_{min}/(16\sqrt{\varepsilon'})$, where $\lambda_{min}$ and $\varepsilon'$ are the minimum wavelength over the band (8.2-12.4 GHz) and highest expected value of the real part of the permittivity (2.5) of the specimen, respectively. Additional improvements in runtime were achieved by leveraging the symmetry of the problem geometry. A perfect electric conductor (PEC) boundary condition was applied at the center height of the computational volume (i.e., $y=0$, along the $x$-$z$ plane) and, similarly, a perfect magnetic conductor (PMC) boundary condition was used at the center width of the computational volume (i.e., $x=0$, along the $y$-$z$ plane). The use of these boundary conditions, illustrated in Figs. 3(c)-3(f), reduces the simulation time by reducing the computational volume by a factor of 4.

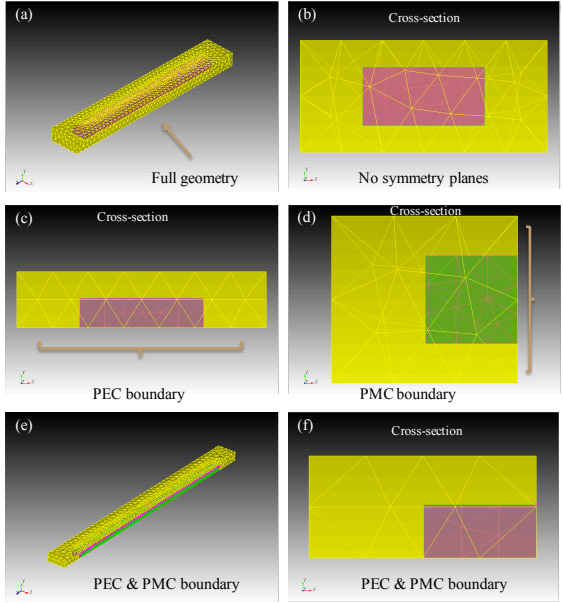

FIG. 3. The model geometry of the specimen inserted into the waveguide with an adaptive tetrahedral mesh. (a) Full geometry without symmetry planes. Cross-sectional view of the (b) full geometry without symmetry planes, (c) geometry with only a PEC boundary at $y=0$, along the $x$-$z$ plane, (d) geometry with only a PMC boundary at $x=0$, along the $y$-$z$ plane. (e) Oblique view of the geometry with both PEC and PMC boundaries and the (f) corresponding cross-sectional view.

Excitation was provided at the port with the $TE_{10}$ mode, and a standard eigenmode matching procedure was applied at the port for termination. In a similar fashion to the experimental measurements described in the measurement stage, empty WG simulations are used as a 'fixture compensation' term to mitigate phase error in the FEM model due to grid dispersion. This calibrated and compensated simulation of $S_{21}$ is then fit to the experimental $S_{21}$ from the second-stage. The fitness function generates a figure of merit to score each simulation, and therefore each complex permittivity value tested for the unknown material. These values are then used to direct the genetic algorithm-based solver to minimize the difference between the measured and simulated values of $S_{21}$.

## III. SENSITIVITY ANALYSIS

The success and applicability of this technique depends on the sensitivity of $S_{21}$ to small changes in the material properties, particularly the loss tangent of the material under test. A two-part sensitivity analysis was performed to determine how the length of the sample influences the sensitivity of the technique and how sample position uncertainty within the waveguide impacts overall measurement uncertainty.

The sensitivity analysis of the partially filled X-band WG CEM method for extracting the permittivity was performed by a series of simulations of $S_{21}$ at 8, 10, and 12 GHz. A baseline material with a relatively low dielectric loss permittivity value ($\varepsilon(\omega)^*=3-j0.003$) was selected to analyze the impacts of slight variations in the real and imaginary parts of the permittivity, Figs. 4(a)-4(b). In order to investigate the sensitivity of the measurement technique to changes in the loss factor ($\varepsilon(\omega)''$), the imaginary component of the permittivity was varied by a factor of ½ and 2, $\varepsilon(\omega)^*=3-j0.0015$ and $\varepsilon(\omega)^*=3-j0.006$, Fig. 4(b), respectively. In a similar fashion, the real part of the permittivity ($\varepsilon(\omega)'$) was varied by ±1%, while maintaining $\tan\delta$ as a constant value ($1\times10^{-3}$), such that $\varepsilon(\omega)^*=3.03-j0.00303$ and $\varepsilon(\omega)^*=2.97-j0.00297$, respectively, Fig. 4(a). Simulations were performed for different lengths ($l_s=2.54$, 5.08, 7.62, 10.16, 12.7, and 15.24 cm) at these complex permittivity values. The specimen cross-section of 1.143 cm x 0.508 cm ($w_s$ x $h_s$), schematically illustrated in Fig. 2(c), was held constant for these simulations. The sensitivity to changes in the real ($\varepsilon(\omega)'$) and imaginary ($\varepsilon(\omega)''$) component of the permittivity can then be directly compared to the sensitivity to spatial variations of 127 $\mu$m (5 mils) of the specimen's position within the WG along the $x$- and $y$-axis (Table 1). Deviations along the $z$-axis do not significantly influence the $S_{21}$ measurement, because the fundamental mode of the WG does not vary in regards to intensity along the propagation direction.

Increasing the length ($l_s$) of the specimen along the propagation direction ($z$) in the X-band WG allows for an increased interaction of the specimen with the electromagnetic field, thus enhancing the sensitivity to the specimen's dissipation factor, shown in Fig. 4. The sensitivity of the measurement to changes in the $\varepsilon(\omega)^*$ needs to be sufficiently higher than the measurement repeatability of the network analyzer and the tolerances to spatial variances of the specimen within the WG. As a conservative threshold we consider the measurement repeatability of the network analyzer to be 0.02 dB and 0.1° in amplitude and phase, respectively. For all cases of $l_s$, the measurement sensitivity to the real part of the permittivity is greater than the measurement repeatability of the network analyzer in phase. For the case that the specimen $l_s=2.54$ cm, the change in 1% of the real part of the permittivity resulted in a ~0.75° change in phase, as shown in Fig. 4(a), over 7 times the measurement repeatability of the network analyzer. However, as expected, the measurement sensitivity requires longer specimen lengths to overcome the anticipated measurement repeatability of the network analyzer in terms of amplitude. For this simulated material, a change of ½ at the $10^{-3}$ for the loss factor for a specimen of $l_s$=12.7 cm results in a change of ~0.03 dB in amplitude of $S_{21}$, as shown in Fig. 4(b), which is greater than the measurement uncertainty of the network analyzer, 0.02 dB. Therefore, a sample of length greater than or equal to $l_s$=12.7 cm is recommended to overcome the conservative network analyzer tolerances used for this analysis. It should be noted that shorting the WG and measuring $S_{11}$ instead of $S_{21}$ should increase the sensitivity of the technique by effectively doubling the electrical length; however, this affect was not studied in detail and goes beyond the scope of this work. Furthermore, this would add an additional parasitic contribution into the uncertainty analysis due to the spatial dependence of the specimen along the $z$-axis, although this could potentially be reconciled by positioning the specimen against the short.

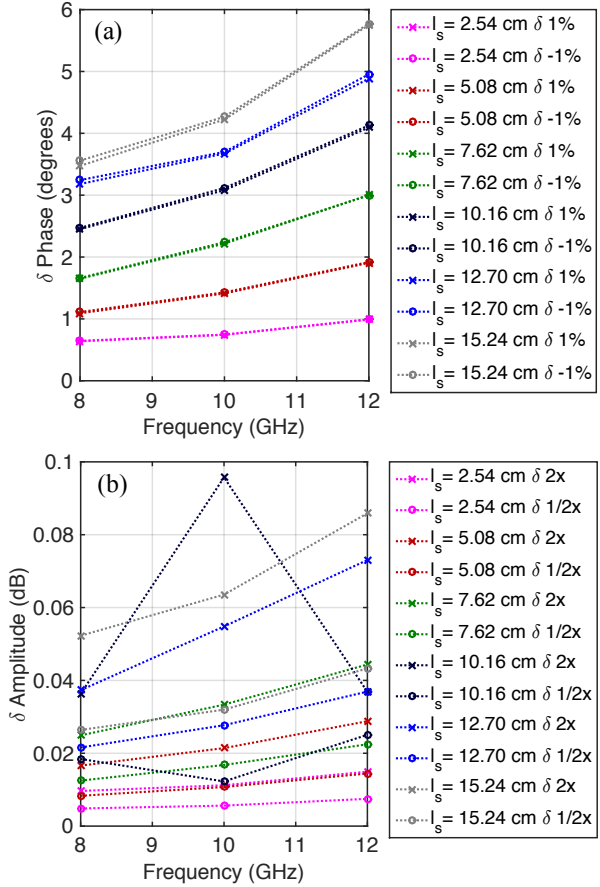

FIG. 4. Sensitivity analysis of the (a) phase and (b) amplitude to the length of the specimen ($l_s$) inserted into the X-band waveguide. (a) The real part of the permittivity ($\varepsilon(\omega)'$) was varied by ±1%. (b) The imaginary component of the permittivity ($\varepsilon(\omega)''$) was varied by a factor of 1/2 and 2. This was performed for $l_s$ =2.54-15.24 cm, in 2.54 cm increments.

TABLE I. Sensitivity analysis of the amplitude (dB) and phase (degrees) to 127 $\mu m$ (5 mil) spatial deviations of the specimen inside the WG for various lengths of specimen.

| Length (cm) | $x \pm 127\ \mu m$ (dB/degrees) | $y \pm 127\ \mu m$ (dB/degrees) |
| --- | --- | --- |
| 2.54 | 0.0005/0.0555 | 0.0014/0.1199 |
| 5.08 | 0.0004/0.0800 | 0.0018/0.1876 |
| 7.62 | 0.0012/0.1172 | 0.0033/0.2925 |
| 10.16 | 0.0002/0.1532 | 0.0050/0.3909 |
| 12.70 | 0.0012/0.1736 | 0.0027/0.4614 |
| 15.24 | 0.0005/0.2031 | 0.0012/0.5271 |

The sensitivity analysis shows that the method presented in this work is robust in terms spatial tolerance; deviations of 127 $\mu m$ along the $x$- or $y$-axis result in changes to $S_{21}$ that are significantly smaller than the change due to a small change in material properties, as shown in Table 1. This scale of deviation can be accounted for in the simulations after accurately measuring the dimensions of the sample holder. For most cases the sensitivity to changes in the permittivity is orders of magnitude greater than changes due to 127 $\mu m$ misalignments of the specimen's position in the WG.

## IV. MEASUREMENT RESULTS

The method presented in this work is validated by measurements in a WR90 WG system, operating from 8.2-12.4 GHz. The vector network analyzer used for the measurements was an Agilent Technologies Vector Network Analyzer Model E8362C. The WR90 WG system was calibrated using the Agilent on-analyzer TRL calibration and a calibration kit provided by the WG manufacturer. This Agilent calibration uses a 10-term error correction model.

As a proof-of-concept, a well-known low dielectric loss specimen (polypropylene) was used for these studies. Based on the results of the sensitivity analysis in the previous section, a specimen of $l_s$=12.7 cm was used for these measurements to ensure we are above the conservative threshold of the measurement uncertainty of the network analyzer. The WR90 WG measurement was also experimentally validated by comparing the resultant material properties with those from a FBS measurement and in literature[9].

Following the procedure in the extraction methodology section, the dimensions of the WG and specimen were collected, Fig. 5(a), to ensure an accurate FEM-model for the extraction procedure. The specimen was measured to be $w_s$=1.176 cm, $h_s$=0.521 cm, $l_s$=12.675 cm and the WR90 had the dimension of $w_{wg}$=2.286 cm, $h_{wg}$=1.143 cm, $l_{wg}$=35.56 cm. After the dimensions were collected for the specimen and WG, the WR90 was TRL calibrated, Fig. 5(b). The specimen was inserted into the WR90 WG, partially filling the cross-sectional area (~25%), as shown in Fig. 5(c). As the final component of the procedure, an empty WG measurement of $S_{21}$ was acquired as a reference and an $S_{21}$ measurement was performed with the sample inserted into the WR90 WG, as shown in Fig. 5(d).

The experimentally acquired $S_{21}$ was used as the target value for the simulations, which took into consideration the proper dimensions from the previous step. The FEM simulation then performed a frequency-by-frequency sweep. At each frequency, a genetic algorithm performed an optimization in order to fit the $S_{21}$ values of the model to the measured values, shown in Fig. 6. It is observed that the $S_{21}$ values are strongly decoupled such that the phase and amplitude correlate to $\varepsilon(\omega)'$ and $\varepsilon(\omega)''$, respectively. However, the amplitude of $S_{21}$ is modulated, Fig. 6(a). These Fabry-Perot resonances formed within the specimen are due to multiple reflections from the front and back interface of the specimen caused by the change of permittivity at the boundaries. This effect couples the amplitude of $S_{21}$ to $\varepsilon(\omega)'$ and $\varepsilon(\omega)''$, where the resonant positions and free-spectral range are dependent on $\varepsilon(\omega)'$ by $2l_s\sqrt{\varepsilon'}\cos\alpha$ and $\lambda_0^2/2l_s\sqrt{\varepsilon'}\cos\alpha$, where $\lambda_0$ is the central free space wavelength of the nearest constructive mode and $\alpha$ is the angle of incidence from the normal of the specimen interface along the *x-y* plane, respectively. On the order of thousands of simulations were performed per frequency to achieve the optimum fit of the amplitude and phase of $S_{21}$, shown in Fig. 6, with an upper bound to the search space set at $\varepsilon(\omega)^*=5-j5$ for the genetic algorithm.

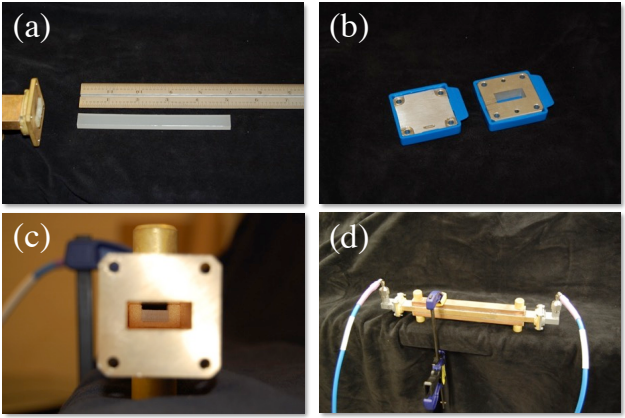

FIG. 5. (a) Polypropylene specimen with a length of ~12.7 cm, (b) electrical short and $\lambda_0/4$ section used for calibration, (c) specimen inserted into the waveguide, partially filling the cross-section, and (d) specimen under test.

The resultant measured real and imaginary permittivity values are shown in Figs. 7(a)-7(b), respectively. For comparison, FBS measurements are shown in Fig. 7 in order to validate the X-band WR90 WG measurement. The FBS measurements were performed on the sheet of polypropylene that was used to machine the material specimen used in the WR90 WG measurements, Fig. 5(a). The two measurement methods are in good agreement in regards to the real part of the permittivity, $\varepsilon(\omega)'\sim 2.25$ for the WG extraction and $\varepsilon(\omega)'\sim 2.28$ for the FBS, Fig. 7(a), which is in good agreement with values reported in the literature, $\varepsilon(\omega)'\sim 2.26\pm 0.04$, from a resonant measurement method[9] at 9.4 GHz. However, the FBS measurement for the imaginary component of the permittivity ($\varepsilon(\omega)''$) is slightly modulated from 0-0.004, and is very sensitive to the sample thickness used in the FBS extraction. The WG measurement provided a significantly more stable result for the imaginary component, $1\times 10^{-5}<\varepsilon(\omega)''<2.6\times 10^{-3}$, Fig. 7(b), which is expected for a non-dispersive specimen. The extracted $\varepsilon(\omega)''$ results for the polypropylene specimen demonstrate the capability of the CEM X-band WR90 WG method and is in good agreement with values reported in the literature of $\varepsilon(\omega)''\approx 1\times 10^{-3}$ at 9.4 GHz by a resonant measurement method[9].

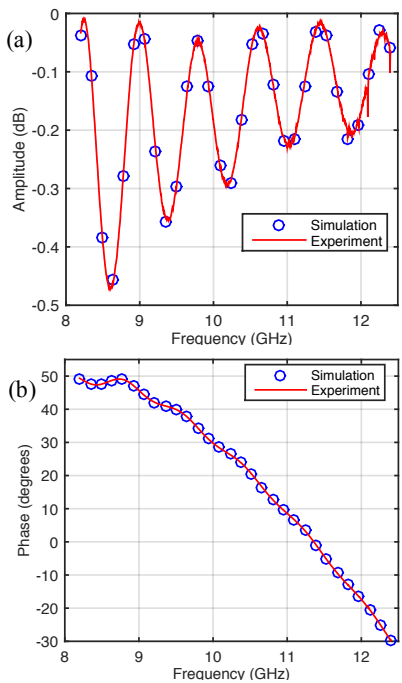

FIG. 6. Simulated and Experimental X-band waveguide measurements of the transmitted scattering parameter ($S_{21}$) (a) amplitude in dB and (b) phase in degrees. These measurements are with a low loss tangent material specimen inserted into the waveguide.

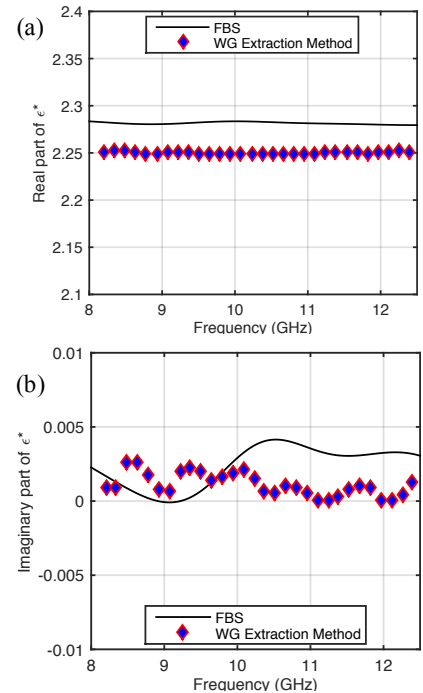

FIG. 7. Focused beam system (FBS) and X-band waveguide (WG) extractions of the (a) real part of the permittivity and (b) imaginary part of the polypropylene's permittivity spectrum.

## V. CONCLUSIONS

In this work, an X-band WG (WR90) measurement method that permits the characterization of the complex permittivity for low dielectric loss tangent material specimen was presented. The extraction method relies on computational electromagnetic (CEM) simulations coupled with a genetic algorithm that performs an optimization scheme which fits the experimental measurement to the simulated transmitted scattering parameter $S_{21}$. It was demonstrated that the sensitivity of the technique is enhanced by increasing the length of the specimen along the direction of propagation in the WG ($z$-axis), while maintaining electrically small dimensions in the cross-section ($x$-$y$ plane). We demonstrated the ability to characterize material specimen of polypropylene with a low dielectric loss tangent value ($<1\times10^{-3}$). This X-band WR90 WG method was validated with good agreement by a traditional free-space focused-beam system measurement of the polypropylene sheet that was used to fabricate the specimen. This technique provides the material measurement community with the ability to accurately extract material properties of low-loss material specimen and could be expanded to an array of materials types (i.e., anisotropic, magnetic, ceramics, high-index) and could easily be adapted for high temperature measurements or other waveguide bands.

The authors would like to thank Daniel L. Faircloth of IERUS Technologies, Inc., for fruitful and stimulating discussions on this work.